\documentclass[preprint,aps]{revtex4}

\usepackage{amsmath}
\usepackage{amssymb}
\usepackage{amsfonts}
\usepackage{epsfig}

\def\b#1{#1}

\def\C{\mathcal{C}}
\def\D{\mathcal{D}}

\def\G{\mathcal{G}}

\def\Alpha{\mathrm{A}}
\def\Epsilon{\mathrm{E}}

\def\d{\mathrm{d}}
\def\H{\mathcal{H}}

\def\Alpha{\mathrm{A}}

\def\C{\mathcal{C}}

\def\d{\mathrm{d}}

\def\G{\mathcal{G}}

\def\H{\mathcal{H}}

\begin{document}

\title{New `phase' of quantum gravity}

\author{Charles H.-T. Wang}
\email{c.wang@abdn.ac.uk}
\affiliation{Centre for Applied Dynamics Research,
School of Engineering and Physical Sciences,
University of
Aberdeen, King's College, Aberdeen AB24 3UE, UK}
\affiliation{Rutherford Appleton Laboratory, Chilton, Didcot, Oxon
OX11 0QX, UK}

\begin{abstract}
The emergence of loop quantum gravity over the past two decades
has stimulated a great resurgence of interest in unifying general
relativity and quantum mechanics. Amongst a number of appealing
features of this approach are the intuitive picture of quantum
geometry using spin networks and powerful mathematical tools from
gauge field theory. However, the present form of loop quantum
gravity suffers from a quantum ambiguity, due to the presence of a
free (Barbero-Immirzi) parameter. Following recent progress on the
conformal decomposition of gravitational fields, we present a new
phase space for general relativity. In addition to spin-gauge
symmetry, the new phase space also incorporates conformal symmetry
making the description parameter free. The Barbero-Immirzi
ambiguity is shown to occur only if the conformal symmetry is
gauge-fixed prior to quantization. By withholding its full
symmetries, the new phase space offers a
promising platform for the future development of loop quantum
gravity. This paper aims to provide an exposition, at a reduced
technical level, of the above theoretical advances and their
background developments. Further details are referred to cited
references.
\end{abstract}

\maketitle

\section{Introduction: quantization of gravity}

Unification is a very cherished concept in theoretical physics. A
classical example is Maxwell's unified electromagnetic theory.
With the birth of special relativity, Einstein further integrated
electromagnetism with the principle of inertia. The work of Dirac
successfully brought together special relativity and quantum
mechanics. A key common feature in all these
achievements is that they rely on no additional physical
constants, spacetime dimensions or material contents other than
already available experimental facts. Nonetheless, these unified
theories lead to experimentally testable predictions: The
electromagnetic wave predicted by Maxwell propagates at the speed
of light calculated from the known permittivity and permeability
coefficients. Special relativity preserves the speed of light as a
universal constant valid in any inertial frames. Dirac's
relativistic quantum theory of electrons predicts intrinsic spin
in units of the measured Planck constant.

In the above examples, it is the new {\em face} of the same
physical entity that the new theory portraits. Such theoretical
advance demands tremendous insights into the interplay
between the physical descriptions to be amalgamated. In
this spirit, Einstein found an elegant reconciliation of Newton's
gravity and special relativity using general relativity,
preserving the spacetime dimensions, gravitational constant and
speed of light. What's radically new is the different
interpretation of gravity in terms of spacetime curvature. Later
efforts to merge general relativity and quantum mechanics have
unfortunately encountered stiff and persistent resistance, and
have given rise to serious doubts on established physical
principles and observed shape of the Universe. Countless ingenious
ideas have been put forward as possible solutions, involving extra
dimensions, super symmetries, strings, branes or combinations of
them. (For a review, see \cite{Carlip2001}.) While these
intellectual yet experimentally unverified hypotheses may well
lead to the ultimate theory of quantum gravity and indeed the
ultimate theory of everything, the search for quantum gravity
without the above postulated structures or
contents of the Universe has continued -- the direction this paper
shall focus on.

\section{Dirac's quantization of constrained Hamiltonian systems}

Dirac expressed strong faith in canonical quantization
\cite{Dirac1964} that has been so successfully applied to quantize
elementary particles and force fields coupling them together,
gravity being an exception. An intriguing aspect of the force
fields in nature is that they all have one kind of symmetry or
another. These are called gauge symmetries and hence the fields
are called gauge fields. Each gauge field has an associated gauge
symmetry group, the simplest example is the U(1) gauge group
for Maxwell's electromagnetic potential field.
Gravity as described by general relativity has symmetries inherent
from the general covariance. Physically this means that there is
no preferred reference frame, either inertial or accelerating.
However, there are at least two reasons for gravity to be so hard
to quantize as a gauge field. First, the theory is highly
nonlinear, and secondly the gauge transformation involves the
change of time itself which is usually fixed in a quantum
evolution. Dirac launched a systematic programme to investigate
the canonical quantization of systems with general gauge
symmetries. He found that the redundancy of
the degrees of freedom due to gauge symmetries can be treated in
an extended Hamiltonian formalism with constraints in a way
similar to the Lagrange multiplier method. The analogue of the
Lagrange constraint function is the constraint
function depending on the canonical coordinates. Such systems are
called constrained Hamiltonian systems. For example, a Hamiltonian
system having a finite dimensional phase space with coordinates
$(q^k,p_k)$ $(k=1,2,\cdots)$ subject to constraints
$\chi_m(q^k,p_k)$ $(m=1,2,\cdots)$ has a Hamiltonian function of
the form:
\begin{equation}
\label{constrHamil}
H(q^k,p_k,\lambda^m)
=
h(q^k,p_k) + \lambda^m \chi_m(q^k,p_k)
\end{equation}
where $\lambda^m$ are Lagrange multipliers.
Einstein's summation convention for repeated indices
is implied throughout this paper.

This extended Hamiltonian structure is so general that the
redundancy does not even have to be of gauge origin. Nonetheless,
the constraints of gauge origin have very simple interpretations
and properties. The time evolution of a system with gauge
symmetries includes gauge transformations generated by the
constraint terms in the Hamiltonian. The constraint plays the role
of the canonical generator of gauge symmetry. The collection of
these constraints can be shown to be algebraically closed. Their
respective Lagrange multipliers in can be arbitrarily specified
reflecting the gauge freedom. In terms of Dirac's terminology,
these constraints are called first class constraints. Other
constraints not having the above properties are not due to gauge
symmetries and are called second class
constraints. When quantizing a system, first class constraints
become quantum constraint operators that annihilate the physical
quantum state. Due to certain consistency
conditions, second class constraints cannot be treated this way
and should in principle be eliminated before quantization.

\section{Quantization ambiguities due to gauge-fixing}
\label{sec:gaugefix}

Given a Hamiltonian system with only first class constraints, i.e.
with any possible second class constraints eliminated, one can in
principle follow Dirac's quantization procedure as described by
treating all constraints on the same footing. On the other hand,
one might try and eliminate some of the constraints to end up with
a Hamiltonian system of fewer degrees of freedom as well as less
constraints. In doing so, some kind of gauge-fixing must be
introduced to remove the respective redundancy in the original
system. As a result, we have a reduced phase space. An interesting
question to ask is whether or not the quantization of the
gauge-fixed system yields an equivalent quantum system as the
quantization of the original system. Not surprisingly, the answer
is `no' in general, as the classical gauge-fixing condition may
violate the uncertainty principle associated with the quantum
variables from the original phase space. Consequently, when a
system is gauge-fixed with respect to a constraint, but in two
different ways, two inequivalent quantizations may arise. This
means that not all states from one quantization can be mapped to
states from the other quantization unitarily. This is an important
source of quantum ambiguity. To avoid it, Dirac's method of
quantization should be performed without gauge fixing any first
class constraints that would lead to a quantum ambiguity.

\section{Geometrodynamics: building spacetime by evolving space}

As a continuum, spacetime described by special relativity has no
absolute time which features in Newtonian dynamics. The Galilean
inertial frames sharing the same time measurement are replaced by
the Lorentz inertial frames, each having their own time. The
underlying geometry is Minkowskian. General relativity goes
further by assuming only local Lorentz frames. In order to
interpret gravity as spacetime curvature, the pseudo-Riemannian
geometry is invoked. Consequently, global Lorentz frames are no
longer available and general spacetime coordinates must be used.
Spacetime now appears to be an elastic continuum that can be
deformed due to massive objects like stars and admits propagating
disturbances, i.e gravitational waves. Indeed, general relativity
can be shown to be the simplest metric theory of gravity to
possess general covariance. The related equivalence principle has
been experimentally verified to a few parts in a trillion~\cite{EvotWash1994}.

\begin{figure}[!h]
\begin{center}
\includegraphics[width=0.6\linewidth]{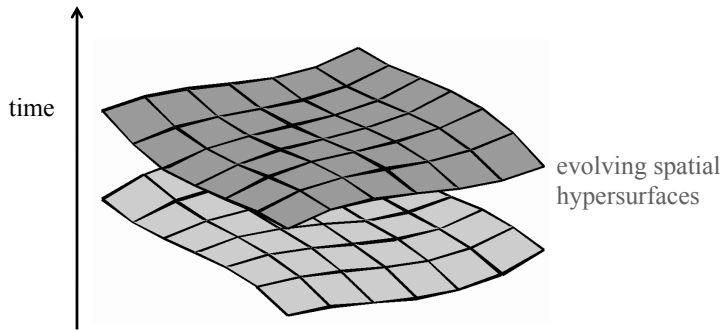}
\caption{{\bf Geometrodynamics} describes spacetime as
a result of the canonical evolution
of a spatial hypersurface with respect to a coordinate time.}
\label{fig:geometrodynamics}
\end{center}
\end{figure}

However, for the purpose of canonically quantizing gravity using
Dirac's prescription, general covariance has to be formally
broken. A somewhat artificial time coordinate must be chosen in
order to define evolution. See Fig.~\ref{fig:geometrodynamics}.
Nonetheless, one hopes that the general covariance will emerge via
the time-slicing independent evolution of spacetime dynamics as
per Dirac's canonical formulation.

\begin{figure}[!h]
\begin{center}
\includegraphics[width=0.5\linewidth]{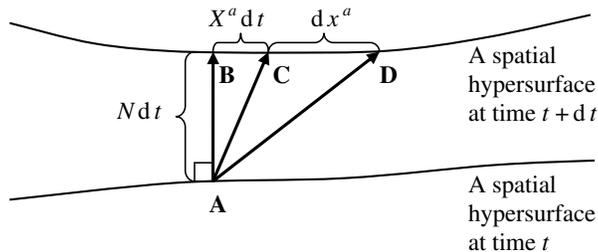}
\caption{\textbf{The effects of the lapse function and shift vector}
are illustrated by considering two constant time hypersurfaces
intersecting event A $(x^a,t)$ and event D $(x^a+ \d x^a,t+\d t)$.
The proper separation between these two events can be  understood intuitively by
considering first the proper time between event A and event B
$(x^a- X^a \d t,t+\d t)$ followed by considering the proper distance between event B and event
C $(x^a,t+\d t)$ and finally the proper distance between events C and D. These considerations lead to
the line element expression given in Eq.~\eqref{ADMmetric}.}
\label{fig:lapse_shift}
\end{center}
\end{figure}

Such a space-time split has been developed by Arnowitt, Deser and
Misner (ADM) \cite{ADM1962, DeWitt1967}. Here, we shall use lower case Greek
letters, e.g. $\alpha, \beta$, to denote spacetime indices ranging
from 0 to 3, and use lower case Latin letters, e.g. $a, b$, to
denote spatial indices ranging from 1 to 3. Starting from
arbitrary spacetime coordinates $(x^\alpha) = (t=x^0, x^a)$, the
spacetime metric $g_{\alpha\beta}$ with metric signature
$(-,+,+,+)$ is decomposed into the time-time component $g_{00}$,
space-time components $g_{0a}$ and space-space components
components $g_{ab}$ accordingly. The space-space components have a
immediate interpretation as they constitute the 3-metric with
metric signature $(+,+,+)$ on the spatial hypersurface with a
constant coordinate time $t$. To interpret the time-time and
space-time metric components, ADM introduced the lapse
function $N=N(x^c,t)$ and the spatial shift vector $X^a=X^a(x^c,t)$
and wrote the squared space time line element in a `$3+1$' fashion
as
\begin{equation}\label{ADMmetric}
\d s^2 = -N^2 \d t^2 + g_{a b} (\d x^a + X^a \d t)(\d x^b + X^b \d t) .
\end{equation}
Equivalently, the spacetime metric and its space-time decomposition can be
written as
\begin{equation}\label{}
\left(g_{\alpha\beta}\right)
=
\left(%
\begin{array}{c@{\;\;\;}c}
  g_{00} & g_{0a} \\
  g_{0a} & g_{ab} \\
\end{array}%
\right)
=
\left(%
\begin{array}{c@{\;\;\;}c}
  -N^2+X^c X_c & X_a \\
  X_a & g_{ab} \\
\end{array}%
\right)
\end{equation}
which shows the clear correspondence between $(N, X^a)$ and $(g_{00},g_{0a})$.
From the expression~\eqref{ADMmetric}, we see the following.
Consider two events $(x^a,t)$ and $(x^a,t+\d t)$
belonging to two nearby spatial
hypersurfaces with coordinate times $t$ and $t+\d t$.
With respect to these hypersurfaces the squared proper time
and length
separations between these two events are given, respectively, by
$N^2\d t^2$ and $g_{a b} (X^a \d t)(X^b \d t)$. This justifies the above
names for $N$ and $X^a$. (See Fig.~\ref{fig:lapse_shift}).

It is well-known that Einstein's field equations can be generated
by varying the Einstein-Hilbert action with respect to the
spacetime metric $g_{\alpha\beta}$~\cite{MTW1973}. The passage
to canonical formulation is via the Legendre
transformation of the Einstein-Hilbert Lagrangian with respect to
the coordinate time derivatives of the spatial metric $g_{ab}$.
Since the time derivatives of the lapse function $N$ and shift
vector $X^a$ are absent in the Einstein-Hilbert Lagrangian, they
become Lagrange multipliers. This leads to the ADM Hamiltonian for
gravity of the following form~\cite{DeWitt1967}:
\begin{equation}
\label{HADM}
H
=
\int\left(N \H + X^a \D_a\right) \d^3 x
\end{equation}
where $p^{ab}$ is the conjugate moment of the metric $g_{ab}$, $\H
= \H[g_{ab},p^{ab}]$ the Hamiltonian constraint and $\D =
\D[g_{ab},p^{ab}]$ the diffeomorphism (or, historically, momentum)
constraint. This Hamiltonian has a natural continuum extension of
Dirac's constrained Hamiltonian of the form \eqref{constrHamil}.
However, the gravitational Hamiltonian \eqref{HADM} constants of
constraint terms only and is therefore `totally constrained'. This
reflects the general covariance of the theory where no spacetime
coordinates are preferred. Generally speaking, the constraint $\D$
generates diffeomorphisms (translations) of ADM's canonical
variables $(g_{ab},p^{ab})$ on the spatial hypersurface while $\H$
generates their time evolution normal to the spatial hypersurface.
Both $\H$ and $\D_a$ are first class and both $N$ and $X^a$ are
arbitrarily specifiable.

The quantization follows formally from Dirac by turning
the classical constraint equations
$N=0$ and $X^a=0$ into the following quantum
constraint equations~\cite{DeWitt1967}
\begin{eqnarray}
\label{weq1}
\hat\H\Psi &=& 0 \\
\label{weq2}
\hat\D_a\Psi &=& 0
\end{eqnarray}
for the quantum state $\Psi=\Psi[g_{a b}]$,
where the quantum operators are obtained from the substitution
$p^{ab}\rightarrow -i\delta/\delta g_{a b}$ in terms of the
functional derivative.

The quantum operator $\hat\D_a$ now generates the
diffeomorphism of the quantum state $\Psi$. Therefore, the quantum constraint
\eqref{weq2} implies that $\Psi$ is diffeomorphism invariant and hence depends on
the spatial geometry instead of the metric.
The quantum Hamiltonian constraint \eqref{weq1} is called the
Wheeler-DeWitt equation and
generates quantum evolution with respect to some
geometric time to be isolated from the spatial geometry.
The above quantization of gravity is conceptually appealing.
However, the lack of suitable functional analytic techniques
means that this approach can at best stay at a formal level.
This difficulty has motivated the developments of alternative
canonical formulations of general relativity in the hope that
a naturally preferred phase space will be found in which the quantization
of gravity is fully implementable.

\section{Spin-gauge variables of gravity}
\label{sec:spingauge}

At present, the most promising approach to canonical general
relativity is based on the use of spin connection variables
that allows general relativity to be reformulated into a
Yang-Mills like gauge field theory. Powerful background
independent quantum field theoretical techniques may then be
invoked and adapted for quantum
gravity~\cite{AshtekarLewandowski2004}. Ashtekar discovered a
particular set of spin connection variables in which the
gravitational constraints take very simple polynomial
forms~\cite{Ashtekar1986,Ashtekar1987}. However, in order to yield real
physical observables, certain reality conditions must be satisfied
but their implementation became problematic.
To resolve this issue,
Barbero put forward
an alternative set of spin connection variables~\cite{Barbero1995}
based on the real spin gauge group SU$(2)$
at the expense of losing polynomiality of the Hamiltonian constraint.
At the first sight, the resulting Hamiltonian constraint appears to be too
complicated to be quantized. Fortunately, the difficulty can be overcome
by a regularization scheme developed by Thiemann~\cite{Thiemann1996,Thiemann1998}.
However, a further issue is a  free parameter
being introduced in
Barbero's construction of the  real spin connection variables, pointed out by Immirzi~\cite{Immirzi1997}.
It is called the Barbero-Immirzi parameter and will result in a
one-parameter ambiguity in the subsequent quantization.
Nonetheless, an avenue has been opened to embrace the above-mentioned
technical advantages by quantizing general relativity as a Yang-Mills
like gauge theory.

\begin{figure}[!h]
\begin{center}
\includegraphics[width=0.6\linewidth]{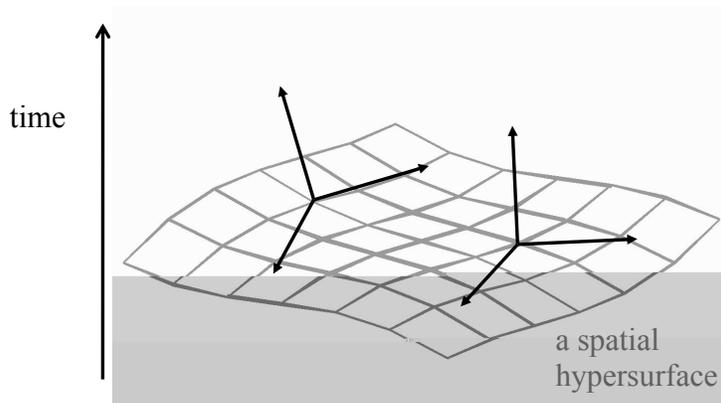}
\caption{{\bf Triads} can be used to specify the 3-metric over a spatial hypersurface.}
\label{fig:triad}
\end{center}
\end{figure}

The formulation can be briefly outlined as follows. Over the spatial hypersurface,
a set of orthonormal vector fields $e^a_i$, called a triad, is introduced.
We use Latin indices staring from $i$ with the range $i,j,\cdots=1,2,3$
to label a member vector of the triad. In terms of the triad, the
contravariant spatial metric is simply
\begin{equation}\label{}
g^{ab} = e^a_i  e^b_i .
\end{equation}
There is an obvious redundancy
in mapping triads to metrics due to an arbitrary 3-dimensional rotation of the triad.
The idea is to replace the metric by the triad as gravitational variables
having a spin-gauge symmetry
with SU(2) as the gauge group. See Fig.~\ref{fig:triad}.
Furthermore, the general SU(2) spin connection $\b{A}^i_a$
may be employed to complete a canonical transformation.
The result is that, for any positive (Barbero-Immirzi)
parameter $\beta$, a rescaled densitized triad $\b{E}^a_i$
can be defined as
\begin{equation}\label{dentriad}
\b{E}^a_i = \beta^{-1} g^{1/2} e^a_i
\end{equation}
where $g = \det g_{ab}$.
Using this,
the following transformation
\begin{equation}\label{}
(g_{ab}, p^{ab}) \rightarrow (\b{A}^i_a, \b{E}^a_i)
\end{equation}
can be shown to be canonical~\cite{Barbero1995, Wang2005c}. Here
$\b{E}^a_i$ is regarded as the momentum of the spin connection
$\b{A}^i_a$ and hence $\b{E}^a_i$ has been made to carry density
weight one in \eqref{dentriad}. It is in complete analogy with the
`electrical field' of the standard SU(2) Yang-Mills gauge theory
where the index $i$ labels a base element in the associated su(2)
Lie algebra.

Just as in Maxwell's U(1) gauge theory and Yang-Mills' SU(2) gauge
theory, here the gravitational analogue of the electrical fields
$\b{E}^a_i$ also satisfies the `Gauss law', i.e. the Gauss
constraint equation $\G_k = 0$ where
\begin{equation}\label{Gk}
\G_k := \b{D}_a \b{E}^a_k
\end{equation}
is called the Gauss constraint and $\b{D}_a$ is the covariant derivative
associated with the connection $\b{A}^i_a$.
This constraint compensates the redundancy in the variables
$(\b{A}^i_a, \b{E}^a_i)$ due to the
spin gauge.

In the spin gauge variables $(\b{A}^i_a, \b{E}^a_i)$, all of the
Hamiltonian constraint $\H$, diffeomorphism constraint $\D$ and
Gauss constraint $\G_k$ are first class and they enter into the
gravitational Hamiltonian according to:

\begin{equation}
H
=
\int\left(N \H + X^a \D_a + Y^k \G_k\right) \d^3 x
\end{equation}
with additional Lagrange multipliers $Y^k=Y^k(x^a,t)$.

\begin{figure}[!h]

\begin{center}
\includegraphics[width=0.4\linewidth]{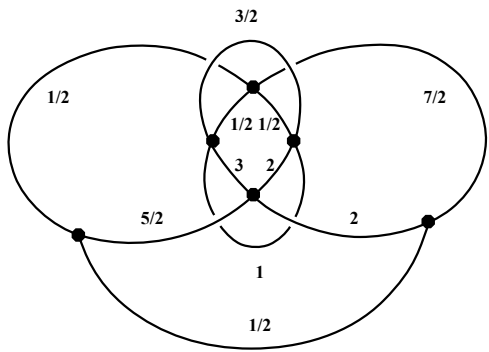}
\caption{{\bf A specimen spin network} consisting of 6 nodes and
11 links. Here 2 nodes are connected to 3 links and 4 nodes are
connected to 4 links. Each link is labelled with a half integer as
a spin quantum number. Each node has also an `intertwiner' quantum
number which is not shown.} \label{fig:sn}

\end{center}
\end{figure}

The Dirac quantization of gravity in these
canonical variables
follows formally as
\begin{eqnarray}
\label{beq1}
\hat\H\Psi &=& 0 \\
\label{beq2}
\hat\D_a\Psi &=& 0 \\
\label{beq3}
\hat\G_k\Psi &=& 0
\end{eqnarray}
where $\Psi=\Psi[\b{A}^i_a]$. Compared with the quantization in
the ADM variables using \eqref{weq1} and \eqref{weq2}, we have now
an additional quantum Gaussian constraint \eqref{beq3}. The
quantization task appears to be more complicated. However, as
stated earlier in this section, the justification for the
spin-gauge formulation is to tap into powerful techniques for
gauge field theories. Specifically, the Gauss constraint
\eqref{beq3} can be solved exactly for Wilson loop states
constructed from the holonomies of the spin connection
$\b{A}^i_a$~\cite{RovelliSmolin1988}, and hence the name `loop
quantum gravity'. By extending Penrose's original spin network
concept~\cite{Penrose1971}, Rovelli and Smolin later generalized
the loop states to spin network states that provide further
mathematical advantages including the availability of a complete
orthonormal basis for all spin gauge invariant
states~\cite{RovelliSmolin1995}. See Fig.~\ref{fig:sn}. For recent
reviews, see e.g.
\cite{AshtekarLewandowski2004,Rovelli2004,Thiemann2006}. However,
these exciting developments still do not address the quantum
ambiguity due to the Barbero-Immirzi parameter.

\section{The need for conformal symmetry}

The quantum ambiguity resulting from the above spin-gauge
formulation is due to the different choice of the scaling
parameter $\beta$. Classically, different choices of $\beta$
correspond to different sets of canonical coordinates for general
relativity. These sets are merely related by canonical
transformations and hence describe the same classical physics.
However, they give rise to inequivalent quantum theories as
demonstrated by Rovelli and Thiemann~\cite{RovelliThiemann1998}.
One view on this problem is that an alternative choice of the
enlarged spin-gauge group SO(4,$\Bbb{C}$) could remove the free
parameter $\beta$~\cite{Alexandrov2000,AlexandrovLivine2003}.
However, its implementation has been complicated by certain second
class constraints. Some other authors see the Barbero-Immirzi
parameter as a parity violation parameter in loop quantum
gravity~\cite{PerezRovelli2006,Freidel2005}. A new viewpoint is
based on the observation that the free parameter $\beta$ defines
the scale of the densitized triad $\b{E}^a_i$. This signals a new
conformal gauge symmetry associated with an underlying fundamental
phase space of general relativity. If we work directly with this
phase space and treat the respective constraint, which is ideally
first class, using Dirac's theory of quantization, then the
Barbero-Immirzi ambiguity may be removed. On the other hand, if
the conformal gauge is fixed by choosing a $\beta$ value, then
this value may enter into the resulting quantization. This way, we
may explain the Barbero-Immirzi ambiguity as a quantum ambiguity
due to gauge-fixing at the classical level as discussed in
section~\ref{sec:gaugefix}.

Naturally, this new phase space may be obtained by extending the
phase space of general relativity with conformal symmetry.
Although the need for this symmetry is motivated from the
spin-gauge formulation of gravity, the conformal gauge by itself
is quite independent of the spin gauge. Furthermore, there are
problems in general relativity where the role of conformal gauge
is of primary importance \cite{BMW2006}. It is
therefore useful to consider first a conformally extended phase
space from that of the geometrodynamics, i.e. the ADM phase space.

The problem is intimately related to the true
dynamical degrees of gravity identified as the conformal
three-geometry by York~\cite{York1971,York1972}. This
identification has provided power tools in the analytical initial
value problems as well as numerical integrations of the
gravitational field. Attempts have been made to apply York's
conformal decomposition in quantum gravity, but have been hammered
by the absence of an appropriate Hamiltonian structure.

\begin{figure}[!h]
\begin{center}
\includegraphics[width=0.6\linewidth]{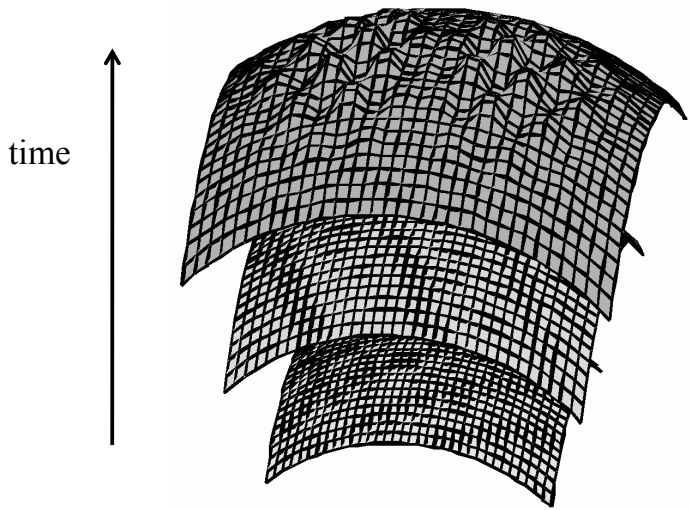}

\caption{{\bf The conformal decomposition of gravity} can be
thought of as separating the scaling part of spatial geometry from
the shearing part. The former describes the expansion of the
Universe, illustrated here as 3 expanding surfaces, while the
latter describes gravitational waves, indicated here as ripples on
the surfaces.}

\label{fig:expansion}
\end{center}
\end{figure}

In a recent paper \cite{Wang2005b}, the canonical evolution of
conformal three-geometry for arbitrary spacetime foliations is
formulated using a new form of Hamiltonian for general relativity.
It is achieved by extending the ADM phase space to that consisting
of York's mean extrinsic curvature time, conformal three-metric
and their momenta. Accordingly, an additional constraint is
introduced, called the conformal constraint.  The complete set of
the conformal, diffeomorphism and Hamiltonian constraints are
shown to be of first class through the explicit construction of
their Poisson brackets. The extended algebra of constraints has as
a subalgebra the Lie algebra for
the conformorphism transformations of the spatial hypersurface.
See Fig.~\ref{fig:expansion}.

The above canonical framework has been
developed into a parameter-free gauge formulation of
general relativity in~\cite{Wang2005c}. (For a review,
see~\cite{Wang2006a}.) The result is a further enlarged set of
first class gravitational constraints consisting of a reduced
Hamiltonian constraint and the canonical generators for spin-gauge
and conformorphism transformations. The
formalism has most recently been simplified into a form
more suitable for quantum implementation, which will form a basis
for the following discussions. Details of
these recent developments will be reported
elsewhere~\cite{Wang2006b,Wang2006c}.

The new starting point is the
canonical transformation of the
gravitational variables of the following form~\cite{Wang2006b}:
\begin{equation}\label{}
(g_{ab}, p^{ab}) \rightarrow (\gamma_{ab}, \pi^{ab}; \phi, \pi)
\end{equation}
where $\gamma_{ab}$ and $\pi^{ab}$ are rescaled from the ADM metric
and momentum
according to
\begin{equation}\label{}
\gamma_{ab} = \phi^{-4} g_{ab},
\quad
\pi^{ab} = \phi^4 p^{ab}
\end{equation}
using an arbitrary positive function $\phi$ as the conformal factor,
with the respective canonical momentum $\pi$. The above construction
obviously involves a local rescaling redundancy. Accordingly,
an additional constraint
\begin{equation}\label{}
\C
=
{\gamma}_{ab}{\pi}^{ab} - \frac14\,\phi\,{\pi} 
\end{equation}
is introduced to offset this redundancy, so that the
number of physical degrees of freedom remains unchanged. It turns
out that the constraint $\C$ is first class and is the canonical
generator of the conformal transformations. It follows from
Dirac's theory of constrained Hamiltonian systems that the
gravitational Hamiltonian in terms of the variables
$(\gamma_{ab},\pi^{ab}; \phi, \pi)$ can be cast in the form:
\begin{equation}
H
=
\int\left(N \H + X^a \D_a + Z \C\right) \d^3 x
\end{equation}
where $Z=Z(x^a,t)$ is a new Lagrange multiplier.
The proof of the above statement and further details
can be found in~\cite{Wang2006b}.

\begin{figure}[!h]
\begin{center}
\includegraphics[width=0.6\linewidth]{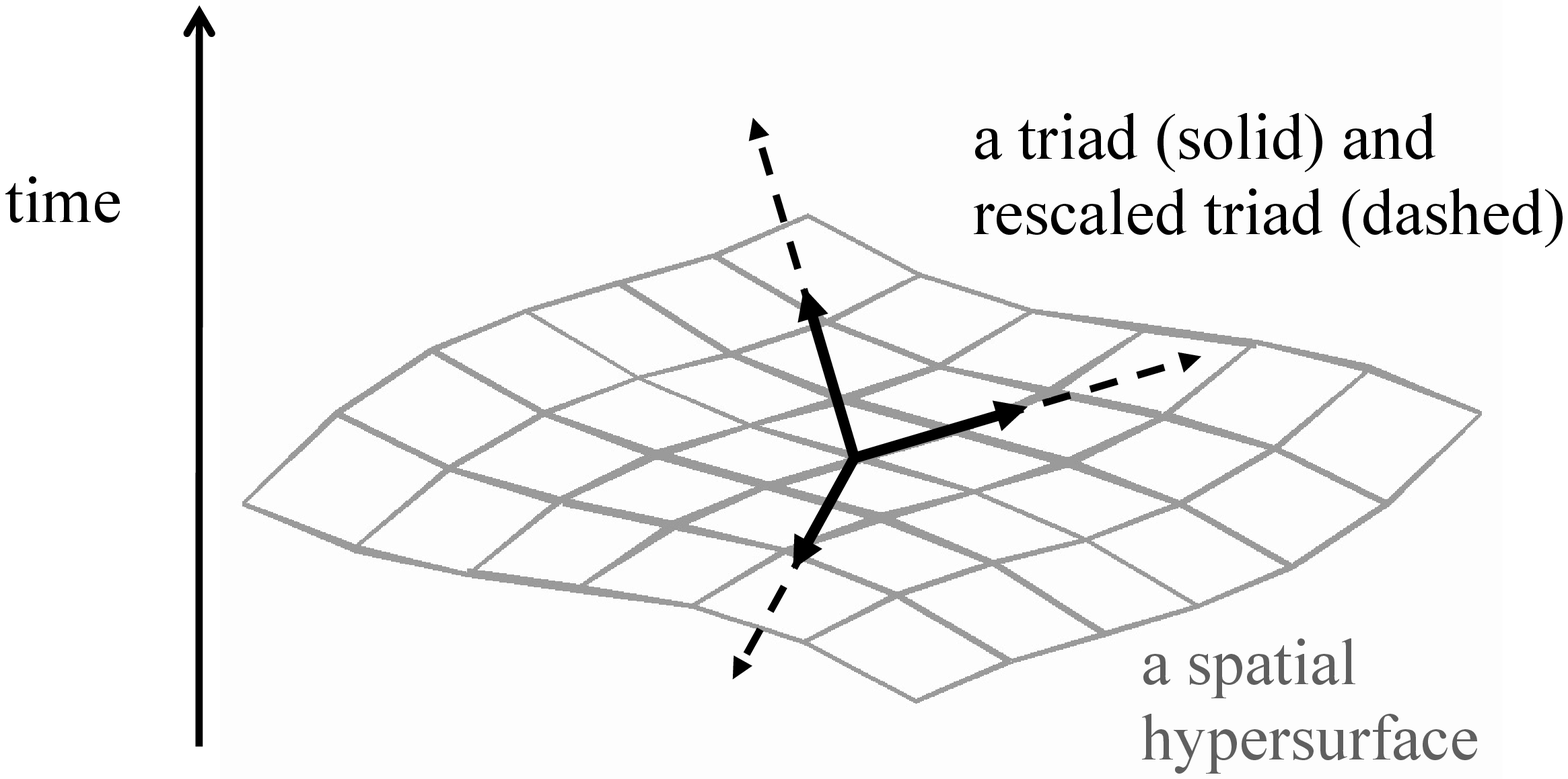}
\caption{{\bf A conformal equivalence class of triads} consists of
triads with the same orientation but different scaling factors.}
\label{fig:triad_rescaled}
\end{center}
\end{figure}

With the availability of the canonical
formulation of general relativity in terms of conformal
equivalence classes of metrics, we may further extend the phase
space to accommodate conformal equivalence classes of triads
(Fig.~\ref{fig:triad_rescaled}) and the corresponding equivalence
classes of SU(2) connections. To this end, introduce the triad
$\epsilon^a_i$ associated with the rescaled metric such that
\begin{equation}\label{}
\gamma^{ab} = \epsilon^a_i  \epsilon^b_i
\end{equation}
By analogy with the densitized triad in \eqref{dentriad}, we
introduce the rescaled densitized triad
by
\begin{equation}\label{}
\Epsilon^a_i
=
\gamma^{1/2} \epsilon^a_i
=
\phi^{-4} g^{1/2} e^a_i .
\end{equation}
Here, we see that
the global scaling parameter in \eqref{dentriad}
has now become a local scaling coefficient according to
$\beta\rightarrow\phi^{1/4}$.
By regarding $\Epsilon^a_i$ as the momentum of the SU(2)
connection ${\Alpha}^i_a$ associated with the rescaled
triad $\epsilon^a_i$, we can complete
the following canonical transformation~\cite{Wang2006c}:
\begin{equation}\label{}
(\gamma_{ab}, \pi^{ab}; \phi, \pi) \rightarrow ({\Alpha}^i_a, {\Epsilon}^a_i; \phi, \pi) .
\end{equation}
We have now redundancies in the variables
$({\Alpha}^i_a,{\Epsilon}^a_i; \phi, \pi)$ due to spin-gauge and conformal
transformations, generated by the Gauss constraint $\G$ and
conformal constraint $\C$ respectively. This leads to our final
form of the gravitational Hamiltonian of the
form~\cite{Wang2006c}:
\begin{equation}
H
=
\int\left(N \H + X^a \D_a + Y^k \G_k + Z \C\right) \d^3 x .
\end{equation}
The detailed construction of the constraints $\H, \D_a, \G_k$ and
$\C$ and the proof of their first class nature are given
in~\cite{Wang2006c}.

The quantization of gravity in the
canonical variables
$({\Alpha}^i_a, {\Epsilon}^a_i; \phi, \pi)$
follows formally as
\begin{eqnarray}
\label{feq1}
\hat\H\Psi &=& 0 \\
\label{feq2}
\hat\D_a\Psi &=& 0 \\
\label{feq3}
\hat\G_k\Psi &=& 0 \\
\label{feq4}
\hat\C\Psi &=& 0
\end{eqnarray}
where $\Psi=\Psi[{\Alpha}^i_a, \phi]$. Here all the constraints
$\H, \D_a, \G_k$ and $\C$ are treated on an equal basis. If we
eliminate the conformal constraint $\C$ at the classical level by
freezing $\phi$ to be a constant, say $\phi=\beta^{1/4}$, then the
phase space reduces immediately to that of the standard spin-gauge
formulation described in section~\ref{sec:spingauge}. Though the
quantization in this reduced phase space using
\eqref{beq1}--\eqref{beq3} has less quantum constraints, the price
to pay is a $\beta$-dependent quantum ambiguity. We see that the
Barbero-Immirzi ambiguity originates from
gauge fixation, as discussed in
section~\ref{sec:gaugefix}. In contrast, the quantization using
\eqref{feq1}--\eqref{feq4} is free from this ambiguity. This
provides strong evidence that the true phase space in which
quantum gravity occurs has the canonical coordinates
$({\Alpha}^i_a, {\Epsilon}^a_i; \phi, \pi)$ and that $\H, \D_a,
\G_k$ and $\C$ constitute the complete set of gravitational
constraints.



\section{Concluding remarks and future vision}

A discussion has been given of recent developments in unifying the
two great theories of modern physics -- general relativity and
quantum mechanics. Inspired by promising progress on loop quantum
gravity, we have reviewed its underlying spin-gauge structure
which enables the powerful loop and indeed the spin network
quantization techniques to conquer the `unquantizable'. However,
the quantum ambiguity due to a free (Barbero-Immirzi) parameter
existing in the present loop quantum gravity suggests that the
theory is not final yet. After all, a motivating rationale for
loop quantum gravity is the very {\em parameter free} approach to
quantum gravity. We take the presence of the Barbero-Immirzi
ambiguity as the indication that the gauge symmetries for
canonical gravity must be further enlarged to incorporate
conformal symmetry, in addition to spin-gauge symmetry. The need
for conformal symmetry originates from the fact that the
Barbero-Immirzi parameter is a scaling parameter. By locally
`gauging' this scaling invariance we obtain an extended phase
space of general relativity with conformal symmetry. Indeed,
starting from this extended phase space we see that the
Barbero-Immirzi ambiguity arises if the conformal symmetry is
gauge fixed prior to quantization, which is equivalent to the
present loop quantum gravity. Therefore, our vision for the new
face of quantum gravity is a fresh new phase space with conformal
and spin-gauge symmetries as the unambiguous basis for unifying
general relativity and quantum mechanics.

\section*{Acknowledgements}
I wish to thank
J. F. Barbero,
R. Bingham,
S. Carlip,
A. E. Fischer,
G. Immirzi,
C. J. Isham,
J. T. Mendon\c{c}a,
N. O'Murchadha
and
J. W. York
for stimulating discussions. The work was supported in part by
the
CCLRC Centre for Fundamental Physics.


\begin{thebibliography}{99}


\bibitem{Carlip2001}
S. Carlip,
Quantum gravity: a progress report,
Rep. Prog. Phys. {\bf 64} 885--942 (2001)


\bibitem{Dirac1964}
P. A. M. Dirac,  Lectures on Quantum Mechanics (Yeshiva
University, New York, 1964)

\bibitem{EvotWash1994}
Y. Su et al,
New tests of the universality of free fall,
Phys. Rev. D {\bf 50}, 3614--3636 (1994)

\bibitem{ADM1962}
R. Arnowitt, S. Deser and C. W. Misner,
Gravitation: An Introduction to Current Research,
ed.  L. Witten (Wiley, New York, 1962)


\bibitem{DeWitt1967}
B. S. DeWitt,
Quantum theory of gravity. I. The canonical theory
Phys. Rev. {\bf 160}, 1113--1834 (1967)


\bibitem{MTW1973}
C. W. Misner, K. S. Thorne and J. A. Wheeler,
Gravitation,
(W. H. Freeman, San Francisco, 1973)


\bibitem{AshtekarLewandowski2004}
A. Ashtekar and J. Lewandowski,
Background independent quantum gravity: a status report,
Classical Quantum Gravity {\bf 21}, R53--R152 (2004)


\bibitem{Ashtekar1986}
A. Ashtekar,
New variables for classical and quantum gravity,
Phys. Rev. Lett. {\bf 57}, 2244--2247 (1986)


\bibitem{Ashtekar1987}
A. Ashtekar,
New Hamiltonian formulation of general relativity,
Phys. Rev. D {\bf 36}, 1587--1602 (1987)

\bibitem{Barbero1995}
J. F. Barbero G.,
Real Ashtekar variables for Lorentzian signature space-times,
Phys. Rev. D \textbf{51}, 5507--5510 (1995)

\bibitem{Thiemann1996}
T. Thiemann,
Anomaly-free formulation of non-perturbative, four-dimensional Lorentzian quantum gravity,
Phys. Lett. B {\bf 380}, 257--264 (1996)

\bibitem{Thiemann1998}
T. Thiemann,
Quantum spin dynamics (QSD),
Classical Quantum Gravity {\bf 15}, 839--873 (1998)



\bibitem{Immirzi1997}
G. Immirzi,
Real and complex connections for canonical gravity,
Classical Quantum Gravity \textbf{14}, L177--L181 (1997)


\bibitem{RovelliSmolin1988}
C. Rovelli and L. Smolin,
Knot theory and quantum gravity,
Phys. Rev. Lett. {\bf61}, 1155--1158 (1988)

\bibitem{Penrose1971}
R. Penrose,
Quantum Theory and Beyond,
ed. T. Bastin
(Cambridge University Press, Cambridge, 1971)


\bibitem{RovelliSmolin1995}
C. Rovelli and L. Smolin,
Spin networks and quantum gravity,
Phys. Rev. D {\bf52}, 5743--5759 (1995)

\bibitem{Rovelli2004}
C. Rovelli,
Quantum gravity
(Cambridge University Press, Cambridge, 2004)

\bibitem{Thiemann2006}
T. Thiemann,
Modern Canonical Quantum General Relativity,
(Cambridge University Press, Cambridge, 2006);
preprint, arXiv:gr-qc/0110034


\bibitem{RovelliThiemann1998}
C. Rovelli and T. Thiemann,
The Immirzi parameter in quantum general relativity,
Phys. Rev. D 57, 1009--1014
(1998)

\bibitem{Alexandrov2000}
S. Yu. Alexandrov,
SO(4,C)-covariant Ashtekar-Barbero gravity and the Immirzi parameter,
Classical Quantum Gravity {\bf17}, 4255--4268 (2000)


\bibitem{AlexandrovLivine2003}
S. Yu. Alexandrov and E. R. Livine,
SU(2) loop quantum gravity seen from covariant theory,
Phys. Rev. D {\bf67}, 044009-15 (2003)


\bibitem{PerezRovelli2006} 

A. Perez and C. Rovelli,
Physical effects of the Immirzi parameter,
Phys. Rev. D {\bf73}, 044013-3 (2006)


\bibitem{Freidel2005} 
L. Freidel, D. Minic and T. Takeuchi,
Quantum gravity, torsion, parity violation, and all that,
Phys. Rev. D {\bf72}, 104002-6 (2005)



\bibitem{BMW2006}
C. H.-T. Wang, R. Bingham, J. T. Mendon\c{c}a,
Quantum gravitational decoherence of matter waves,
Classical Quantum Gravity {\bf 23}, L59--L65 (2006)


\bibitem{York1971}
J. W. York, Gravitational degrees of freedom and the initial-value
problem, Phys. Rev. Lett. {\bf 26}, 1656 (1971)

\bibitem{York1972}
J. W. York,
Role of coformal 3-geometry in the dynamics of gravitation,
Phys. Rev. Lett. {\bf 28}, 1082--1085 (1972)

\bibitem{Wang2005b}
C. H.-T. Wang,
Conformal geometrodynamics: True degrees of freedom in a truly canonical structure,
Phys. Rev. D {\bf 71}, 124026-7 (2005)

\bibitem{Wang2005c}
C. H.-T. Wang,
Unambiguous spin-gauge formulation of canonical general relativity with conformorphism invariance,
Phys. Rev. D {\bf 72}, 087501-4 (2005)

\bibitem{Wang2006a} 
C. H.-T. Wang,
Towards conformal loop quantum gravity,
J. Phys. Conf. Ser. {\bf33}, 285--290 (2006)


\bibitem{Wang2006b}
C. H.-T. Wang,
Conformal decomposition in canonical general relativity,
preprint, arXiv:gr-qc/0603062


\bibitem{Wang2006c}
C. H.-T. Wang,
The conformal factor in the parameter-free construction of spin-gauge variables for gravity,
preprint, arXiv:gr-qc/0603077


\end{thebibliography}
\end{document}